\documentclass{aastex}
\usepackage{spr-astr-addons}
\usepackage{url}

\RequirePackage{color}

\begin{document}

\title{Double cyclic variations in orbital period of the eclipsing cataclysmic variable EX Dra}
\slugcomment{Not to appear in Nonlearned J., 45.}
\shorttitle{The eclipsing dwarf nova EX Dra}
\shortauthors{Han Z. T. et al.}

\author{Zhong-tao Han\altaffilmark{1,2,3}} \and \author{Sheng-bang Qian\altaffilmark{1,2,3}}
\and \author{Irina Voloshina\altaffilmark{4}}
\and \author{Li-Ying Zhu\altaffilmark{1,2,3}}

\altaffiltext{1}{Yunnan Observatories, Chinese Academy of Sciences (CAS), P. O. Box 110, 650216 Kunming, China; zhongtaohan@ynao.ac.cn}
\altaffiltext{2}{Key Laboratory of the Structure and Evolution of Celestial Objects, Chinese Academy of Sciences, P. O. Box 110, 650216 Kunming, China.}
\altaffiltext{3}{University of Chinese Academy of Sciences, Yuquan Road 19\#, Sijingshang Block, 100049 Beijing, China.}
\altaffiltext{4}{Sternberg Astronomical Institute, Moscow State University, Universitetskij prospect 13, Moscow 119992, Russia.}

\begin{abstract}
EX Dra is a long-period eclipsing dwarf nova with $\sim2-3$ mag amplitude outbursts. This star has been monitored photometrically from November, 2009 to March, 2016 and 29 new mid-eclipse times were obtained. By using new data together with the published data, the best fit to the $O-C$ curve indicate that the orbital period of EX Dra have an upward parabolic change while undergoing double-cyclic variations with the periods of 21.4 and 3.99 years, respectively. The upward parabolic change reveals a long-term increase at a rate of $\dot{P}={+7.46}\times10^{-11}{s} {s^{-1}}$. The evolutionary theory of cataclysmic variables (CVs) predicts that, as a CV evolves, the orbital period should be decreasing rather than increasing. Secular increase can be explained as the mass transfer between the secondary and primary or may be just an observed part of a longer cyclic change. Most plausible explanation for the double-cyclic variations is a pair of light travel-time effect via the presence of two companions. Their masses are determined to be $M_{A}sini'_{A}=29.3(\pm0.6) M_{Jup}$ and $M_{B}sini'_{B}=50.8(\pm0.2) M_{Jup}$. When the two companions are coplanar to the orbital plane of the central eclipsing pair, their masses would match to brown dwarfs.
\end{abstract}



\section{Introduction}

To search substellar objects around white dwarf binaries are important for understanding of the interaction between companions and evolved stars \citep[]{2015ApJS..221...17Q}. Recent years have been reported several successful examples for the detection of the planets around the white dwarf binaries such as QS Vir \citep*[]{2010MNRAS.401L..34Q, 2011IAUS..276..495A}, NN Ser \citep{2014MNRAS.437..475M}, HU Aqr \citep{2011MNRAS.414L..16Q}, \citep{2015MNRAS.448.1118G} and RR Cae \citep{2012MNRAS.422L..24Q}. These substellar companions are detected by measuring the variations in the observed mid-eclipse time via the presence of the third body. As the motion of the binary near the common center of mass, the arrival eclipse time will vary periodically. This method was widely used to study other eclipsing binary systems containing a white dwarf and a red dwarf because the components are large differences in radius and luminosity \citep[]{2010MNRAS.407.2362P}.

Cataclysmic variables are semi-detached binaries containing a white dwarf accreting material
from a main-sequence star via Roche lobe overflow \citep[]{1995CVSC...562A..19V}. As one of the subclasses of CVs, dwarf novae show recurrent outbursts with the amplitude of 2-5 mag and short duration about a few days to weeks. Recently, several eclipsing dwarf novae were selected to detect the substellar companions and evolution by analyzing variations in orbital period. One of good examples is V2051 Oph, Qian et al. (2015) reported that it has a giant planet with a mass of $7.3(\pm0.7)M_{Jup}$ and an eccentricity of $e'=0.37$. Moreover, the secular decrease in orbital period of V2051 Oph suggested that magnetic braking may not entirely cease in fully convective stars. EX Dra is a long-period ($P=5.04$ h) dwarf nova with very deep eclipse with 1.5 mag in quiescence. It was discovered in the Hamburger Quasar Survey \citep[]{1989ESASP.296..883B}. Follow-up observations by \citet{1993IAU.165..89B} showed that EX Dra is a deeply eclipsing dwarf nova with an orbital period of just over 5 h. \citet{1997A&A...327..173F} presented spectroscopic and photometric observations, and estimated some basic parameters.
By using photometric observations, \citet{2000MNRAS.316..529B} found that this system has a mass ratio $q={0.72}$ and an inclination $i=85^{\circ}$, and that the $O-C$ diagram showed a periodic oscillation with a period of 4 yr and an amplitude of 1.2 min.
\citet{2003PASP..115.1105S} analyzed the eclipse profile of multi-colour light curves with a parameter-fitting model. They derived a mass ratio $q<0.81$ and an inclination of $i>83^{\circ}$. The revised ephemeris showed a cyclical variation with a period of 5 yr. Recently, the analysis by \citet{2012A&A...539A.153P} given a
period modulation with a period of 21 yr and an amplitude of 2.5 min.

In present paper, we use new eclipse timings coupled with the old data to analyse the $O-C$ diagram of EX Dra. Our results indicated that both there are two possible brown dwarfs orbiting EX Dra, and this star may be undergoing a peculiarly evolutionary stage.

\section{CCD photometric observations and new mid-eclipse times}

We started to monitor EX Dra since 09, November 2009 by using the 0.6-m reflecting telescope attached an Andor DV436 2K CCD camera at the Yunnan Observatories(YNAO). Later, this star was monitored with the 85-cm telescope mounted an Anor DW436 1K CCD camera at the XingLong station of the National Astronomical Observatories and the 2.4-m telescope at the Lijiang observational station in YNAO .
Since the May of 2014, EX Dra was continuously observed with CCD photometer on 50-cm (Apogee Alta U8300 with 528 x 512 pixels) and 60-cm (Apogee 47, field 1024 x 1024 pixels) telescopes of Sternberg Astronomical Institute Crimean Station in R-band.
Four light curves of EX Dra in quiescence are displayed in Figure 1. The phases were computed by using the linear ephemeris,
\begin{equation}
Min.I=HJD2456065.154955+0.209937316\times{E},
\end{equation}
where HJD2456065.154955 is the initial epoch from our mid-eclipse times listed in Table 1, and 0.209937316 d is the orbital period from \citet{2012A&A...539A.153P}.
The most obvious features in Figure 1 are that the light curves in quiescence exhibit double eclipse rarely and strong orbital hump. In addition, the shape and brightness are variable with time. This can be explained as the change of mass transfer rate and the unstability of accretion disc. The egress of white dwarf can be seen clearly in the light curves. For comparison, two profiles during outburst are shown in Figure 2. These curves are V-shaped and symmetric, indicating an axisymmetrical brightness distribution in the accretion disc at maximum. The orbital hump disappears during outburst.

\begin{figure}[!h]
\begin{center}
\includegraphics[width=1.0\columnwidth]{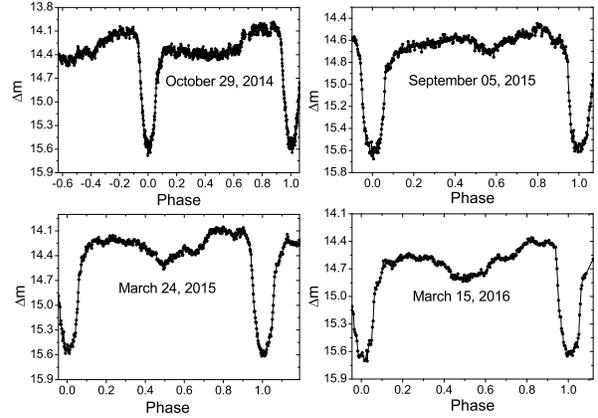}
\caption{Four eclipsing light curves of EX Dra observed by using the 60cm telescope in Sternberg Astronomical Institute Crimean Station.}
\end{center}
\end{figure}

\begin{figure}[!h]
\begin{center}
\includegraphics[width=1.0\columnwidth]{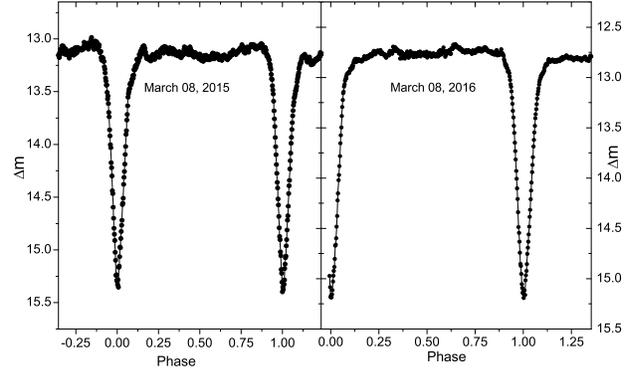}
\caption{Two eclipse profiles of EX Dra during outbursts obtained with 60cm telescope in Sternberg Astronomical Institute Crimean Station on 2015 March 08 and 2016 March 08, respectively.}
\end{center}
\end{figure}

By adopting the same method from \citet{2012A&A...539A.153P}, the 29 new mid-eclipse times during quiescence were obtained and listed in Table 1.
We only used the light curves during quiescence to determine the mid-eclipse times. The reason also has been discussed in \citet{2012A&A...539A.153P}. This consistency with previous published eclipse timings highly increases reliability of the following analysis.
We also computed the eclipse width of the white dwarf as $\Delta{\tau}=0.0230(1)$ days, which is very close to previous studies.
The uncertainty in determining mid-egress times depends on the time resolution and signal to noise ratio. We estimate that the error of mid-eclipse times is about quarter of the integration time. The reason is that the errors of mid-egress and mid-ingress times are about half of the time resolution, and the mid-eclipse times are average value of the mid-ingress and mid-egress times. Therefore, the errors of mid-eclipse times were determined as the combination of the errors of mid-egress and mid-ingress by using the error propagation function.
All new mid-eclipse times have been converted into the $BJD$ system and are listed in the second column of Table 1, corresponding to errors are also given in fifth column.
The exposure time for each mid-eclipse times was listed in sixth column.
The details of
the used filters could be found in seventh column where "R" and "I" refer to R-band and I-band, respectively. "N" indicates that no filters were used. "0.6m", "85cm" and "2.4m" in the eighth column of the table refer to the 0.6m, 85-cm and 2.4-m telescopes in China,  while "50cm" and "60cm" refer to the 50-cm and 60-cm telescopes in Russia.

\section{The changes of the $O-C$ curve of EX Dra}

\citet{2000MNRAS.316..529B} shown a cyclical behaviour in $O-C$ diagram with a period of 4 years and a amplitude of 1.18 min. Follow-up studies by \citet{2003PASP..115.1105S} pointed out the period and amplitude are about $25\,\%$ bigger than the corresponding values from \citet{2000MNRAS.316..529B}. Recently, \citet{2012A&A...539A.153P} revised ephemeris by adding many mid-eclipse times and found a greater cyclical variation with a period of 21 years and an amplitude of 2.5 min, but a singly sinusoidal ephemeris cannot describe the complex $O-C$ change well. Therefore, it seems that there are two cyclic variations in the $O-C$ diagram.

Combining new data with the old timings from the literature \citep[]{1997A&A...327..173F, 2000MNRAS.316..529B, 2003PASP..115.1105S, 2012A&A...539A.153P}, the latest $O-C$ diagram was obtained (see Figure 3). All $O-C$ values were calculated with the linear ephemeris published by \citet{2012A&A...539A.153P},
\begin{equation}
Min.I=BJD2452474.80513+0.209937316\times{E}, 
\end{equation}
where BJD$2452474.80513$ is the initial epoch. New $O-C$ curve is more complex than meets the eye. Based on previous studies, we suspected the existence of two cyclic variations. To describe the overall trend of the $O-C$ curve well, a quadratic ephemeris is required. Thus, a possible model with an upward parabolic variation and double periodic terms is considered:
\begin{equation}
O-C=\Delta{T_{0}}+\Delta{P_{0}}{E}+\frac{\beta}{2}{E^{2}}+\tau_{A}+\tau_{B}, 
\end{equation}
where $\tau_{A}$ and $\tau_{B}$ are the two cyclic changes.
Our best fit to the $O-C$ diagram by using the Levenberg-Marquardt method shows that both of $\tau_{A}$ and $\tau_{B}$ are strictly periodic, i.e.
\begin{equation}
\tau_{A}=K_{A}\sin(\frac{2\pi}{P_{A}}{E}+\varphi_{A}), 
\end{equation}
and
\begin{equation}
\tau_{B}=K_{B}\sin(\frac{2\pi}{P_{B}}{E}+\varphi_{B}). 
\end{equation}
In general, the eccentricity should be taken into account in the fitting process. However, the eccentricity was close to zero ($e<0.01$) but with a larger error, which is why we set e=0 in the final fit. All fitting parameters and the corresponding values are given in Table 2.
The best-fitting results reveal a secular period increase at a rate of $\dot{P}={+7.46}\times10^{-11}{s} {s^{-1}}$.
In Figure 3, the dashed line in the upper panel refers to the linear period increase and the solid line represents the combination of two cyclic changes and the linear increase. After the long-term increase was subtracted, the superposition of a long (the dashed line) and a short (the solid line) periodic variation are displayed in the middle panel. Following both the linear increase and the two cyclic changes were removed, the residuals are plotted in the lowest panel. The two cyclic variations extracted from the middle panel of Figure 3 are displayed in Figure 4 where the periods of $\tau_{A}$ and $\tau_{B}$ are 21.40 years and 3.99 years and the corresponding amplitudes are 89.6 s and 50.1 s.
The derived period modulations are very close to the previous results detected by \citet{2012A&A...539A.153P} and \citet{2000MNRAS.316..529B}, respectively.

\begin{table*}[!h]
\caption{New CCD mid-eclipse times of EX Dra in quiescence.}
 \begin{center}
 \small
   \begin{tabular}{llcllccc}\hline\hline
Min.(HJD)            & Min.(BJD)         &  E         & $O-C$     & Errors    & Exp.time(s)         & Filters  &Telescopes    \\\hline
2455144.9992         &2455144.9999       &12719	      &0.0021	    &0.0002     & 60                &  R  &0.6m\\
2455746.0491         &2455746.0498       &15582	      &0.0014	    &0.0001     & 40                &  R  &85cm\\
2456065.1550         &2456065.1557       &17102	      &0.0026	    &0.0002     & 60                &  N  &2.4m\\
2456523.0282         &2456523.0290       &19283	      &0.0026	    &0.0001     & 40                &  I  &85cm\\
2456799.3056         &2456799.3063       &20599	      &0.0024     &0.0001     & 40                  &  R  &50cm\\
2456807.2833         &2456807.2841       &20637	      &0.0026     &0.0001     & 40                  &  R  &60cm\\
2456819.4599         &2456819.4607       &20695	      &0.0028     &0.0001     & 40                  &  R  &50cm\\
2456833.3160         &2456833.3167       &20761	      &0.0030     &0.0001     & 40                  &  R  &50cm\\
2456834.3655         &2456834.3663       &20766	      &0.0028     &0.0001     & 40                  &  R  &60cm\\
2456838.3540         &2456838.3548       &20785	      &0.0025     &0.0001     & 40                  &  R  &60cm\\
2456960.3272         &2456960.3280       &21366	      &0.0022     &0.0004     & 75                  &  R  &60cm\\
2456960.5372         &2456960.5379       &21367	      &0.0022     &0.0004     & 75                  &  R  &60cm\\
2457024.1480         &2457024.1488       &21670	      &0.0020     &0.0004     & 75                  &  R  &60cm\\
2457080.2018         &2457080.2026       &21937	      &0.0026     &0.0002     & 60                  &  R  &60cm\\
2457098.2561         &2457098.2569       &22023	      &0.0023     &0.0002     & 60                  &  R  &60cm\\
2457106.2335         &2457106.2342       &22061	      &0.0020     &0.0001     & 40                  &  R  &60cm\\
2457106.4443         &2457106.4451       &22062	      &0.0029     &0.0001     & 40                  &  R  &60cm\\
2457108.3340         &2457108.3347       &22071	      &0.0031     &0.0001     & 40                  &  R  &60cm\\
2457118.4105         &2457118.4112       &22119	      &0.0026     &0.0001     & 40                  &  R  &60cm\\
2457123.2390         &2457123.2398       &22142	      &0.0026     &0.0001     & 40                  &  R  &60cm\\
2457123.4493         &2457123.4501       &22143	      &0.0029     &0.0001     & 40                  &  R  &60cm\\
2457124.4968         &2457124.4976       &22148	      &0.0008     &0.0001     & 40                  &  R  &60cm\\
2457267.2561         &2457267.2569       &22828	      &0.0027     &0.0001     & 40                  &  R  &60cm\\
2457267.4668         &2457267.4676       &22829	      &0.0034     &0.0001     & 40                  &  R  &60cm\\
2457268.3058         &2457268.3066       &22833	      &0.0028     &0.0002     & 60                  &  R  &60cm\\
2457271.2457         &2457271.2465       &22847	      &0.0035     &0.0002     & 60                  &  R  &60cm\\
2457271.4561         &2457271.4568       &22848	      &0.0039     &0.0001     & 40                  &  R  &60cm\\
2457463.3380         &2457463.3388       &23762	      &0.0032     &0.0001     & 40                  &  R  &60cm\\
2457463.5479         &2457463.5487       &23763	      &0.0032     &0.0001     & 40                  &  R  &60cm\\
\hline
\end{tabular}
\end{center}
\end{table*}

\begin{table*}
\caption{Parameters of the best fitting for $O-C$ .}
\label{tab:table2}
\begin{center}
\small
\begin{tabular}{lllllllll}
\hline
\hline
Parameters                                                          &Values                                \\
\hline
Correction on the initial epoch, ${\Delta{T_{0}}}$ (d)           & $-6.0(\pm1.6)\times10^{-4}$          \\
Correction on the initial period, ${\Delta{P_{0}}}$ (d)          & $+3.09(\pm0.28)\times10^{-8}$        \\
Rate of the  linear increase, $\beta$ (d/cycle)                  & $+1.57(\pm0.26)\times10^{-11}$       \\
                                                                                                     \\
Parameters                                                         & Case A   &Case B                                                                                                            \\
Semi-amplitude, $K_{A}$, $K_{B}$ (d)                            & $1.04(\pm0.24)\times10^{-3}$ & $5.80(\pm0.73)\times10^{-4}$ \\
Orbital period, $P_{A}$, $P_{B}$ (yr)                           & $21.40(\pm1.44)$  & $3.99(\pm0.11)$                      \\
The orbital phase, $\varphi_{A}$, $\varphi_{B}$ (deg)              & $11.17(\pm0.86)$  & $-146.57(\pm5.26)$                      \\
\hline
\end{tabular}
\end{center}
\end{table*}

\begin{figure}[!h]
\begin{center}
\includegraphics[width=1.0\columnwidth]{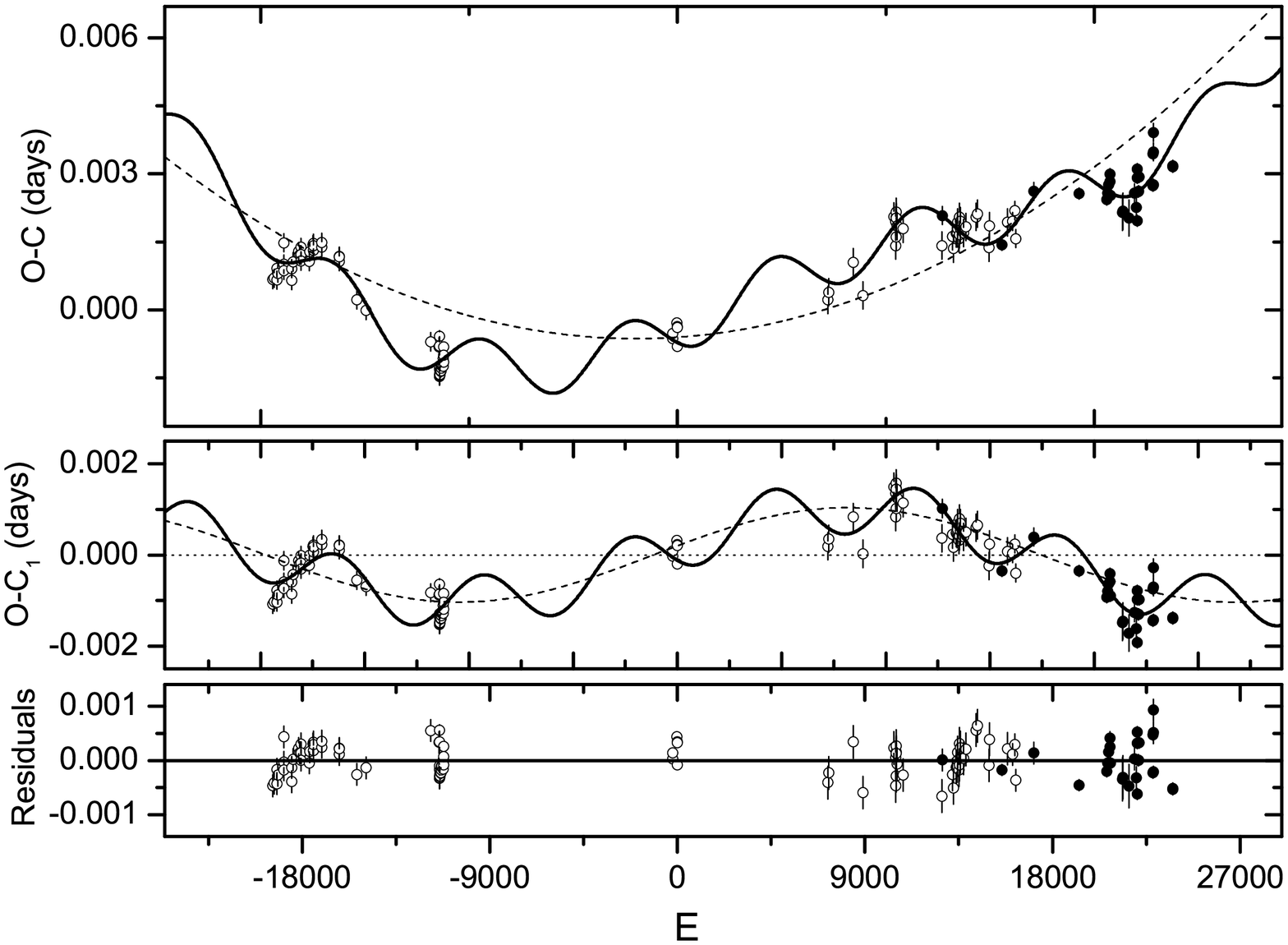}
\caption{$O-C$ diagrams of EX Dra with respect to the double-cyclic variations. The open circles and solid circles denote the data in literature and in our observation, respectively. The solid line in the upper panel refers to a combination of a upward parabolic and two cyclic changes. The dashed line represents only the upward parabolic variation that reveals a continuous increase in the orbital period. When the long-term period increase was subtracted, the superposition of a long (the dashed line) and a short (the solid line) periodic variation are displayed in the middle panel. These variations were removed, the residuals are plotted in the lowest panel.}
\end{center}
\end{figure}

\begin{figure}[!h]
\begin{center}
\includegraphics[width=1.0\columnwidth]{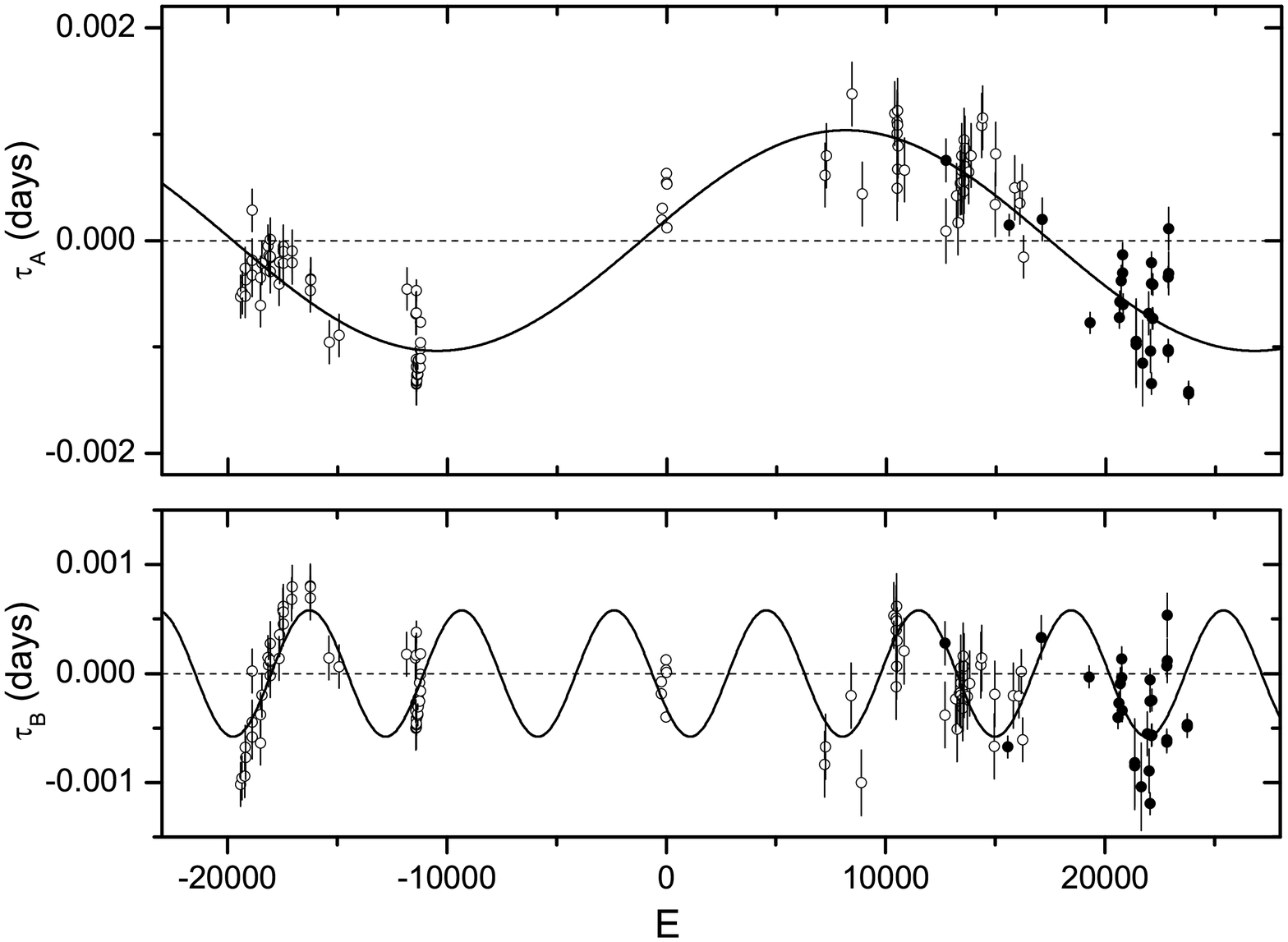}
\caption{The two cyclic variations $\tau_{A}$ and $\tau_{B}$ extracted from the middle panel of Figure 3.}
\end{center}
\end{figure}

\section{Discussion}

The standard model predicts that the evolution of CVs is driven by angular momentum losses(AMLs). The result is that, as a CV evolves, the orbital period decreases.
However, our result show that the period of EX Dra is increasing at a rate of $\dot{P}={+7.46}\times10^{-11}{s} {s^{-1}}$. EX Dra is a long-period ($P=5.04$ h) CV containing a late-type main sequence star overfilling its Roche lobe ($M_{2}\sim0.54M_{\odot}$) and a white dwarf ($M_{1}\sim0.75M_{\odot}$) \citep[]{2000MNRAS.316..529B}, the mass transfer between two components will cause the orbit expansion. Supposing a conservation mass transfer on long time scales and adopting the parameters given by \citet{2000MNRAS.316..529B}, a calculation using the equation \citep[]{1977ARA&A..15..127T}
\begin{equation}
\frac{\dot{P}}{P}=-3\dot{M}_{2}\times(\frac{1}{M_{1}}-\frac{1}{M_{2}}),
\end{equation}
leads to a mass transfer rate of $\dot{M_{2}}=8.34\times10^{-8}M_{\odot} \\ yr^{-1}$. It is alternatively possible that the quadratic term is only a part of a longer cyclic oscillation.

Our results also reveal that there are two cyclic variations in the $O-C$ curve. To interpret cyclic period changes of EX Dra, two main mechanisms are the solar-type magnetic activity cycle in M-type secondary star \citep[]{1992ApJ...385..621A} and the light time travel effect. The Applegate mechanism built on the basis of the conclusion presented by \citet{1989SSRv...50..219H}.
They found that all cool component stars are strictly limited in the spectral types later than F5. Recently, however, a statistical investigation for the cyclic period oscillations has shown that the percentages of cyclic variations for both late-type and early type interacting binaries are very close \citep[]{2010MNRAS.405.1930L}. Thus, the conclusion proposed by \citet{1989SSRv...50..219H} may be not correct, and moreover , \citet{1992ApJ...385..621A} already noted that his model should be revised if the shell becomes a significant fraction of the star's mass ($M_{s}>0.1M_{2}$).
The secondary star in EX Dra is a late-type main sequence star with the spectral type about M1-3/5 \citep[]{2003A&A...404..301R}(update RKcat7.23 version, 2015), this star should have a very deep convective envelope. With the theoretical model calculation and statistics, the shell mass of EX Dra's donor was estimated to be $M_{s}\approx0.15M_{\odot}\approx 0.28M_2$. To explain the cyclic oscillation of the pre-CV NN Ser, moreover, \citet{2006MNRAS.365..287B} by comparing the energies required to cause the observed variation found that NN Ser's secondary star cannot provide enough energy to drive Applegate mechanism. Using the same method for EX Dra, the required energies to produce two cyclic oscillations were calculated and shown in Figure 5. The results show that the required minimum energy in Case A are larger than the total energy radiated in 21.4 yr, and in Case B the required minimum energy are also slightly larger than the total energy radiated in a whole cycle (see Figure 5). Combining the parameters presented by \citet{2000MNRAS.316..529B} with Kepler's third law
\begin{equation}
{P^{2}_{orb}}=\frac{4\pi^{2}a^{3}}{G(M_{1}+M_{2})},
\end{equation}
to yield the orbital separation as $a=1.62R_{\odot}$. Applying $T_{2}=3400K$ for the M2-3 type, the luminosity of the secondary star can be draw as $L_{2}=(\frac{R_{2}}{R_{\odot}})^{2}(\frac{T_{2}}{T_{\odot}})^{4}L_{\odot}$.
Therefore, the Applegate mechanism is difficult to explain the observed cyclic changes.
The most plausible explanation seems to be a pair of light travel time effects via the presence of two companions.

The mass function and the mass of tertiary companions were derived by using the following equation \citep[]{1985ibs..book.....P}:
\begin{equation}
f(m)_A=\frac{4\pi^{2}}{GP_{A}^{2}}(a_{A}^{'}sini^{'}_{A})^{3}=\frac{(M_{A}sini'_{A})^{3}}{(M_{1}+M_{2}+M_{A})^{2}},
\end{equation}
and
\begin{equation}
f(m)_B=\frac{4\pi^{2}}{GP_{B}^{2}}(a_{B}^{'}sini^{'}_{B})^{3}=\frac{(M_{B}sini'_{B})^{3}}{(M_{1}+M_{2}+M_{B})^{2}},
\end{equation}
where $G$ is the gravitational constant, $P_A$ and $P_B$ are the periods of $\tau_{A}$ and $\tau_{B}$, and $a'_{A}sini'_{A}$ and $a'_{B}sini'_{B}$ can be determined by
\begin{equation}
a'_{A}sini'_{A}=K_{A}\times c ,
\end{equation}
and
\begin{equation}
a'_{B}sini'_{B}=K_{B}\times c ,
\end{equation}
$K_A$ and $K_B$ are the semi-amplitude of $\tau_{A}$ and $\tau_{B}$. The results are listed in Table 3. Assuming a random distribution of orbital plane inclinations, the orbital inclination for the companion A (i.e. Case A) is larger than $22^{\circ}.96$, the mass corresponds to $M_{A}\leq0.075M_{\odot}$, it may be a brown dwarf with $74.5\,\%$ and a low-mass star with only $25.5\,\%$ probability. As for companion B (corresponding to Case B), if its orbital inclination is less than $42^{\circ}.39$, the mass is $M_B\geq0.075M_{\odot}$, it may be a brown dwarf with $52.9\,\%$ and low-mass star with $47.1\,\%$. If they are coplanar (i.e. $i'=i=85^{\circ}$) to the orbital plane of the eclipsing pair, their masses would match to two brown dwarfs.

\begin{table*}
\caption{Orbital parameters of the circumbinary substellar objects.}
\begin{center}
\small
\begin{tabular}{lllllllll}
\hline
\hline
Parameters                                                          &Companion A          &Companion B         \\
\hline
Eccentricity, $e_A$ and $e_{B}$                                     & 0                 & 0                \\
Mass function,$f(m)_A$ and $f(m)_B$ $(M_{\odot})$                                  & $1.28(\pm0.63)\times10^{-5}$ &$6.37(\pm0.33)\times10^{-5}$        \\
The companion masses, $M_{A}\sin{i'_A}$ and $M_{B}\sin{i'_B}$ $(M_{Jup})$           & $29.3(\pm0.6)$     & $50.8(\pm0.2)$                   \\
Semi-major axis of the planet, $d_{A}$ and $d_{B}$ ($au,{i'=90^{\circ}}$)       & $9.83(\pm1.21)$     & $3.18(\pm0.11)$                \\
\hline
\end{tabular}
\end{center}
\end{table*}

\begin{figure}[!h]
\begin{center}
\includegraphics[width=1.0\columnwidth]{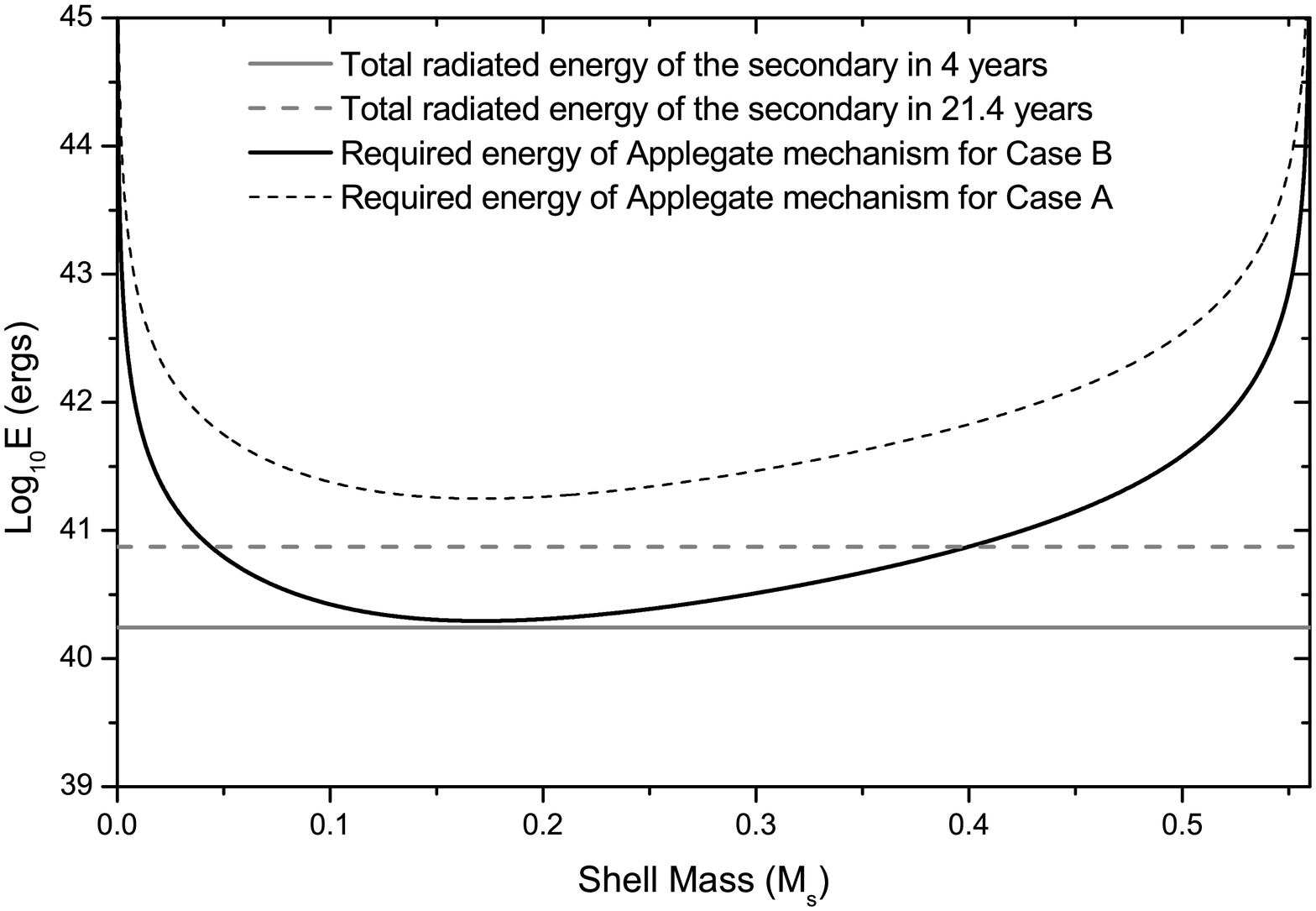}
\caption{Energy required to cause two periodic changes in the $O-C$ diagram by using Applegate's mechanism. $M_s$ refers to the assumed shell mass of the secondary star. The black dashed line denotes the energy required for different shell mass in Case A, and  the black solid line corresponding to Case B. The grey solid line represents the total energy radiates from the secondary in 4 years and the grey dashed line is the total radiant energy of the secondary in 21.4 years.}
\end{center}
\end{figure}

\section{Conclusions}

We have published 29 new mid-eclipse times of EX Dra in quiescence spanning from 2009 to 2016. These mid-eclipse times were used to analyze the orbital period variation.
Besides a secular increase with a rate of $\dot{P}={+7.46}\times10^{-11}{s} {s^{-1}}$, the orbital period also shows the double-cyclic changes. According to the evolutionary theory of CVs, the orbital period should decrease.
If the long-term period increase was explained as the mass transfer from the secondary to primary star, the derived mass transfer rate is $\dot{M_{2}}={8.34}\times10^{-8}M_{\odot}yr^{-1}$.
However, it is possible that the quadratic term may be just a observed part of a longer cyclic oscillation.
For double periodic oscillations, EX Dra's secondary star can not provide enough energy to satisfy the energy requirements of Applegate mechanism. The more acceptable explanation is the existence of a pair of substellar objects around EX Dra. Assuming the circumbinary objects to be in the orbital plane ($i'=i=85^{\circ}$) of the eclipsing pair, they are two brown dwarfs.

The orbital parameters of the substellar objects in Table 3 reveal some interesting features. First, both the orbits are circular. Second, the orbital periods of $21.40(\pm1.44)$ and $3.99(\pm0.11)$ years are nearly the ratio of $5:1$. This implies that the possibility exists for the mean-motion resonance between the two companions and their orbits would be stable.
From the evolutionary perspective, CVs are products of a common envelope (CE) phase \citep[]{2013A&ARv..21...59I}. The circumbinary companions may originate from a large protoplanetary disc or a fragmentation of protostellar disc.
In the former case, the formation process is similar to two hot subdwarf stars HW Vir and AA Dor \citep[]{2000A&A...356..665R, 2009ARA&A..47..211H}, and its description will not be repeated here. In the latter case, the formation process is as follows: the objects will started with the mass a few $M_{Jup}$ and then increase their mass by accreting material from the disc \citep[]{2009A&A...495..201A, 2009MNRAS.392..413S}. For the former formed objects, they would migrate inwards and gain enough mass to become stars \citep[]{2007MNRAS.382L..30S}; for the objects staying in the outer disc region, they could not gain enough mass, and become brown dwarfs \citep[]{2009MNRAS.400.1563S}. The higher-mass objects of the inner region will evolve to progenitor of the post-common envelope binaries. The circumbinary companions formed at the almost same as their hosts and survived the CE phase \citep[]{2014MNRAS.444.1698B}.
Besides, there are also other possibilities, such as the second generation substellar originated in CE event \citep[]{2014A&A...562A..19V}. However, there is only a remote possibility for EX Dra because THE substellar objects have relatively large mass($29.3$ and $50.8$$M_{Jup}$) \citep[]{2014MNRAS.444.1698B}. Moreover, the circular orbits means that they may have existed for a long time-scales before the CE phase. Certainly, in order to confirm our conclusion, further observations are needed in the future.

\acknowledgments
This work is supported by the Chinese Natural Science Foundation (Grant No. 11325315, 11133007, 11573063 and 11611530685), the Strategic Priority Research Program ``The Emergence of Cosmological Structure'' of the Chinese Academy of Sciences (Grant No. XDB09010202) and the Science Foundation of Yunnan Province (Grant No. 2012HC011). This study is also supported by the Russian Foundation for Basic Research (project No. 17-52-53200).
New CCD photometric observations of EX Dra were obtained with the 60cm and the 2.4m telescopes at the Yunnan Observatories, the 85cm telescope in Xinglong Observation base in China and 50cm and 60cm telescopes of Sternberg Astronomical Institute Crimean Station. Finally, we thank the anonymous referee for those helpful comments and suggestions.

\end{document}